\documentclass[12pt,manuscript]{aastex}

\def\BmV0{\mbox{(B-V)$^{\rm o}$}}
\def\VmK0{\mbox{(V-K)$^{\rm o}$}}
\def\MV0{\mbox{M$_{\rm V}^{\rm o}$}}
\def\MV{\mbox{M$_{\rm V}$}}

\def\etal{\mbox{{\it et al. \ }}}

\def\cs22892{{\rm CS~22892-052}}
\def\asec {^{\prime\prime}}
\def\amin {^{\prime}}
\def\fsec {{\rlap.}^{\rm s}}
\def\fasec {{\rlap.}^{\prime \prime}\hskip0.05em}
\def\deg  {^\circ}
\def\about  {$\sim$}

\begin{document}

\title{A RADIO AND X-RAY STUDY OF HISTORICAL SUPERNOVAE IN M83 }

\author{Christopher J. Stockdale\altaffilmark{1}, Larry A. Maddox\altaffilmark{2}, 
John J. Cowan\altaffilmark{2}, Andrea Prestwich\altaffilmark{3}, 
Roy Kilgard\altaffilmark{3}, Stefan Immler\altaffilmark{4}, and Miriam Krauss \altaffilmark{5}
}

\altaffiltext{1}{Department of Physics, Marquette University, PO BOX 1881, Milwaukee, WI, 53201-1881;
christopher.stockdale@marquette.edu}

\altaffiltext{2}{Homer L. Dodge Department of Physics and Astronomy, University of Oklahoma, 440 West Brooks Room 131, Norman, OK 73019; maddox@nhn.ou.edu, cowan@nhn.ou.edu}

\altaffiltext{3}{Harvard-Smithsonian Center for Astrophysics,
60 Garden St, Cambridge, MA, 02138; aprestwich@cfa.harvard.edu, rkilgard@cfa.harvard.edu}

\altaffiltext{4}{ Exploration of the Universe Division,
  X-ray Astrophysics Laboratory, Code 662,
  NASA Goddard Space Flight Center, Greenbelt, MD 20771;
  immler@milkyway.gsfc.nasa.gov}

\altaffiltext{5}{Department of Physics, Massachusetts Institute of Technology, 
Building 6-113,
77 Massachusetts Avenue,
Cambridge, MA, 02139; miriam@space.mit.edu}

\begin{abstract}
We report the results of 15 years of radio observations of the six historical supernovae (SNe) in M83
using the Very Large Array.
We note the near linear decline in radio emission from SN 1957D, a type II SN, which remains a non-thermal radio emitter. The measured flux densities from SNe~1923A and 1950B have flattened as they begin to fade below detectable limits, also type II SNe.
The luminosities for these three SNe are comparable with the radio luminosities of other decades-old SNe
at similar epochs.  SNe~1945B, 1968L, and 1983N were not detected in the most recent observations and these
non-detections are consistent with previous studies.
We report the X-ray non-detections of all six historical SNe using the Chandra X-ray
Observatory, consistent with previous X-ray searches of other decades-old SNe, and low inferred mass
loss rates of the progenitors ($\dot{M} \approx 10^{-8}~M_{\odot}~{\rm yr}^{-1} [v_{\rm w}/10~{\rm km~s}^{-1}]$).
\end{abstract}


\section{Introduction}

M83, an SABc galaxy, is a typical grand design, face-on spiral galaxy with an inclination
of $24\deg$ \citep{tal79}.  It is relatively nearby with distance estimates ranging
 from 3.75 Mpc \citep{dev79} to 8.9 Mpc \citep{san87}. 
We have opted to use the Cepheid-established distance of 4.5 Mpc \citep{thi03}.   
The close proximity and low inclination of M83, along with six optically
discovered supernovae (SNe) make it an
ideal system for studying extragalactic SNe at all
wavelengths.  Here we report on a radio and X-ray study of six
historical SNe.

Using the Very Large Array (VLA),\footnote{The VLA telescope of the National 
Radio Astronomy Observatory is operated by Associated
Universities, Inc. under a cooperative agreement with the National Science 
Foundation.} radio observations have detected emission from four historical SNe, 
SNe~1923A, 1950B,
1957D, and 1983N \citep{cb82,cb85,crb94,spw84}, while
two others, SNe 1945B and 1968L have not been detected in the radio. 
SN 1968L lies within the bright, diffuse, radio
nuclear emission and would not have been detectable unless it was extremely radio bright.
This 20 year study of M83 by Cowan and collaborators has yielded an 
insight into the evolution of the long-term
transient sources in M83, primarily SNe, supernova remnants (SNRs), 
and its nucleus.
\citet{che84} proposes that synchrotron radiation is produced in the region
of interaction between the SN blastwave and the circumstellar shell that 
originated from the prior mass loss of the progenitor star.  In such
models, the emission fades as the density of the circumstellar material 
(CSM) decreases.  \citet{cow84} suggest a minimum of 100 years before 
the radio emission re-brightens as the blastwave begins to encounter 
the interstellar medium, as the source evolves into a SNR.  
\citet{crb94} charted
the evolution of 14 of the brightest radio sources over the course of 
ten years at 6~cm and 20~cm.
We have reanalyzed the \citet{crb94} results and and discovered at 
least five times the number of discrete
radio sources in M83, the complete discussion will be presented
 in Maddox \etal (2005, in preparation).  \citet{imm99} 
detected 37 discrete X-ray sources in M83 with ROSAT observations, 
but no radio emission was detected from the historical SNe.

In this paper  
we report new radio observations of the historical SNe of M83
including  their
current flux densities, spectral indices, and decay indices,
examining how their radio emission has varied during
the time that they have been monitored.  We also give upper limits to
the X-ray flux for the historical SNe.

\section{Observations}
\subsection{Radio}
The new VLA
observations of M83 were made with four observing runs.
In the first two, M83 was observed for 8.7 total hours
on 13 and 15 June 1998, at 20~cm (1.450 GHz) using the
VLA in its BnA configuration (the southern arms in B configuration and
the northern arm in A configuration),
with a maximum baseline of 24 km.
During the second group of observing runs, M83 was
observed on 31 October and 1 November 1998 with the
VLA in its CnB configuration (maximum baseline of \about 11 km)
at 6~cm (4.860 GHz) for a total of 8.9 hours.  
These were done using two 25 MHz bands in dual polarization, split in 8 spectral channels each. 
For the purposes of this study, we are only interested in the continuum
observations of the SNe and have only analyzed the Channel 0, ``pseudo-continuum,'' 
data which utilizes
75\% of the 50 MHz bandwidth in each of
the two orthogonal circular polarizations. The phase calibrator was
J1316-336 with 3C286 used to set the flux density scale for J1316-336.
In each observation, the pointing center was (J2000) $13^{\mathrm{h}}37^{\mathrm{m}}00\fsec22$ and
$-29\deg 52\amin 04\fasec5$.

\begin{figure}
\epsscale{0.8}
\plotone{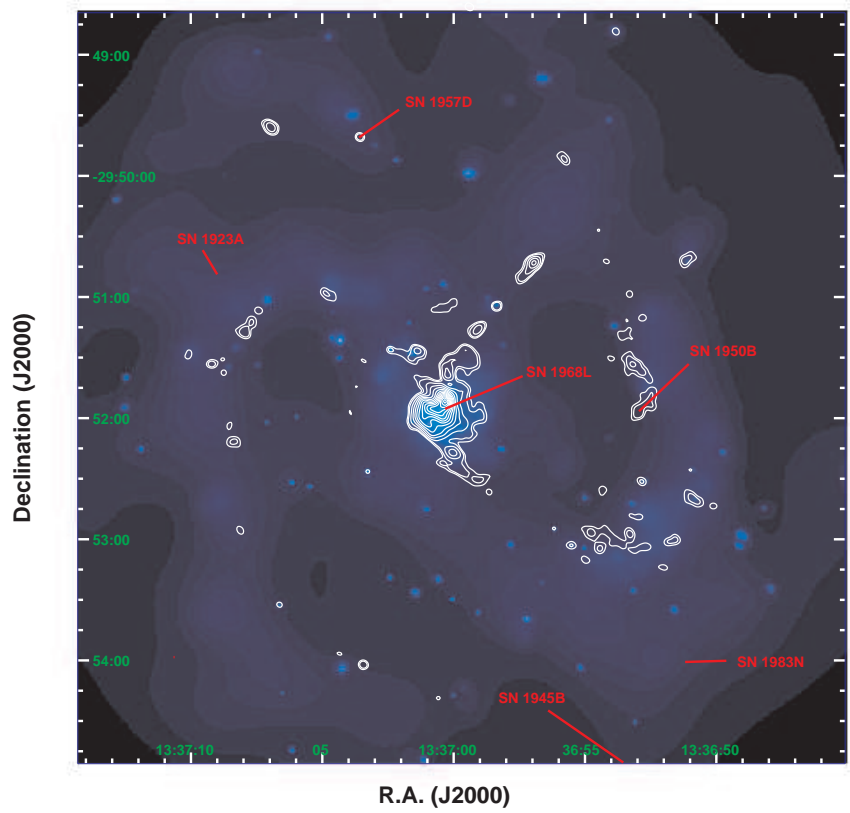}
\caption{
Radio contours (20~cm; 1998) overlaying the soft X-ray (0.3 -- 2~keV; 2001) image of M83.
\label{ovrly}}
\end{figure}

Data were Fourier transformed and deconvolved using the CLEAN algorithm
as implemented in the Astronomical Image Processing System (AIPS) routine SCIMG.
The data were weighted using
Brigg's robustness parameter of $0$, which minimizes the
point-spread function while maximizing sensitivity.
In addition, SCIMG implements a self-calibration algorithm
to reduce the sidelobes from the brightest emission in the nucleus.  The use
of this algorithm allowed us to reduce the noise and to detect many faint sources.
Uncertainties in the peak intensities are reported as the
rms noise from the observations.  At 20~cm, the beam size is
$3\fasec 67 \times 3\fasec 13$, p.a. = $63.58\deg$,
and the rms noise is 0.040 mJy beam$^{-1}$.
At 6~cm, the beam size is
$3\fasec 52 \times 3\fasec 16$, p.a. = $273.0\deg$,
and the rms noise is 0.037 mJy beam$^{-1}$.  
The results of our analyses of the 1998 positions and peak flux densities
are presented in Table~1 and
Figure \ref{ovrly}.  
All flux density measurements were made with the AIPS
routine IMFIT using a two component model,
a 2-d Gaussian component, and a linear sloping background component,
which was the model used by \citet{crb94}.
The errors reported with the flux measurements and the positions were also determined by IMFIT. 
The previous work of \citet{crb94} and \citet{cb85} were also 
re-reduced and re-imaged with the same techniques used for the 1998 data.  
To aid in the identification and modeling of point sources, the 
20~cm data taken in 1983 and 1992 were deconvolved using restoring beams similar in shape to the 
1998 data (rather than the highly elliptical beams derived by SCIMG).
A more detailed description of the data reduction of the earlier observations will presented in 
Maddox \etal (2005, in preparation).

\subsection{X-ray}
M83 has been observed twice with Chandra using ACIS-S.  The
first observation was
a 50ks exposure  in April
2000 (ObsId 793) and the second a 10ks observation in  September 2001 (ObsId 2064).  These
observations have been reduced as described by \citet{kil05}.
No X-ray sources were detected at the positions of historical
supernovae in either exposure \citep{sor02,kil05}.  
Upper limits to the 0.3 -- 8~keV
X-ray luminosity are given
in Table~1.  The upper limits were derived from the longer
observation (ObsId 793)
by integrating the number of valid X-ray counts in a 5-arcsecond aperture
centered on the position of the SNe.  Typically 10-20 X-ray counts were
obtained at each position.  Since most of the counts are likely to be
diffuse background intrinsic to the galaxy, the luminosities in
Table~1 are hard upper limits. 

\section{Results \& Discussion}

\subsection{Radio Emission from Historical Supernovae}

With the VLA observations in 1998,
we detect radio emission from the sites of
SNe~1923A, 1950B, and 1957D
at 20~cm and 6~cm, coincident, within the error limits, of
the sources detected by \citet{crb94} and \citet{eck98}.
No radio emission was detected from SNe 1945B, 1968L, and 1983N, and the implications of these non-detections 
are discussed later at the end of this
Section.
The measured flux densities at 20~cm and 6~cm (presented in Table 1) indicate a decrease 
from 1992 to 1998  for SN 1957D, but no
significant changes for SNe~1923A and 1950B.  SN 1957D has faded by 58\% at
both observed wavelengths from the prior observations in the early 1990's.

\begin{figure}
\plotone{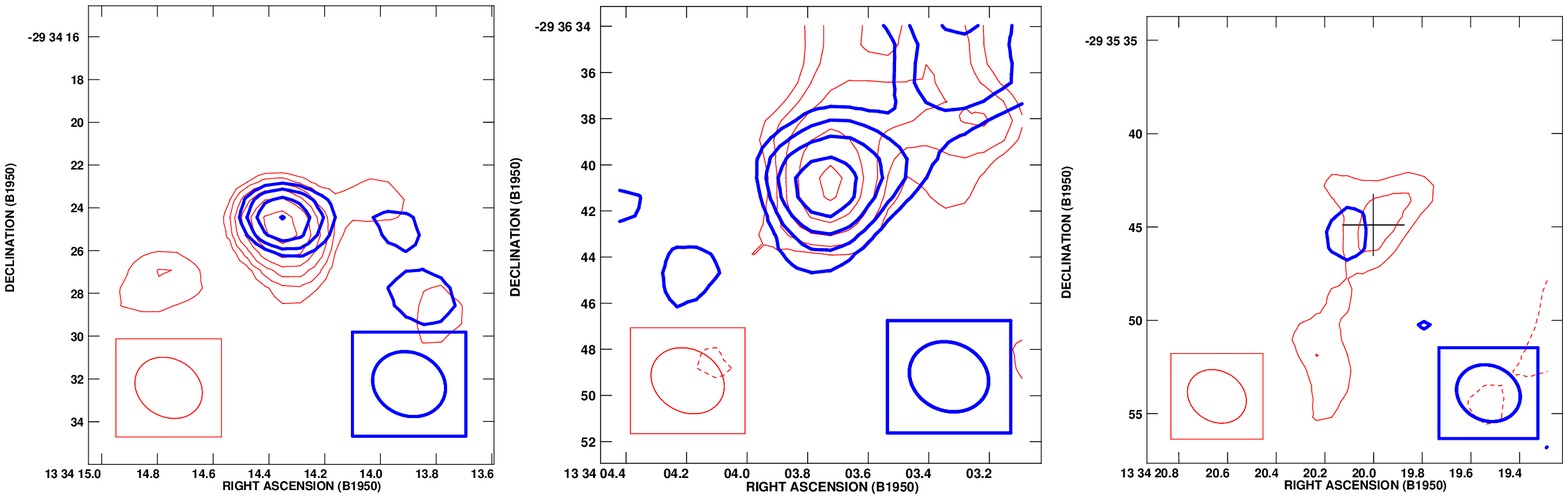}
\caption[Radio contour images at 20~cm and 6~cm of                              
SNe~1957D, 1950B, and 1923A.]{Radio contour images at 20~cm (in red) and 6~cm   
(in blue) of                                                                    
SNe~1957D, 1950B, and 1923A (from left to right).                               
Contour levels at both wavelengths                                              
are $-$0.12, 0.12, 0.17, 0.24, 0.34, 0.48, 0.68, and 0.96 mJy                     
beam$^{-1}$.                                                                    
At 20~cm, the beam size is shown in the lower left,                             
and at 6~cm, the beam size is shown in the lower right. \label{spixfig}}

\end{figure}

As indicated in Table 1 (and shown in Figures \ref{ovrly}, \ref{spixfig}),
our new observations indicate that the spectral indices ($\alpha$; $S \propto \nu^{\alpha}$),
for both SNe~1923A and 1950B are flat.  Our re-analysis of the results of \citet{eck98} and \citet{crb94} indicate that spectral 
indices of both SN~1923A and SN~1950B is unchanged between the early and late 1990s.  
Assuming both were correctly identified as SNe, the typical model for radio emission predicts a steadily decreasing, 
non-thermal emission \citep{wei02}.
In the case of these SNe, their radio emission is likely 
already at or below the level
of the thermal emission from an intervening HII region along the line-of-sight
\citep{mon97}.
SN~1957D however, remains non-thermal and continues to fade at both wavelengths,
as predicted by the models of \citet{wei86,wei90,wei02}.
The evolution
of these three SNe is consistent with the theoretical models
of \citet{wei86,wei90,wei02} and \citet{mon97} for
 radio SNe in which the 6~cm light-curve peaks before the 20~cm emission.

\begin{figure}
\plotone{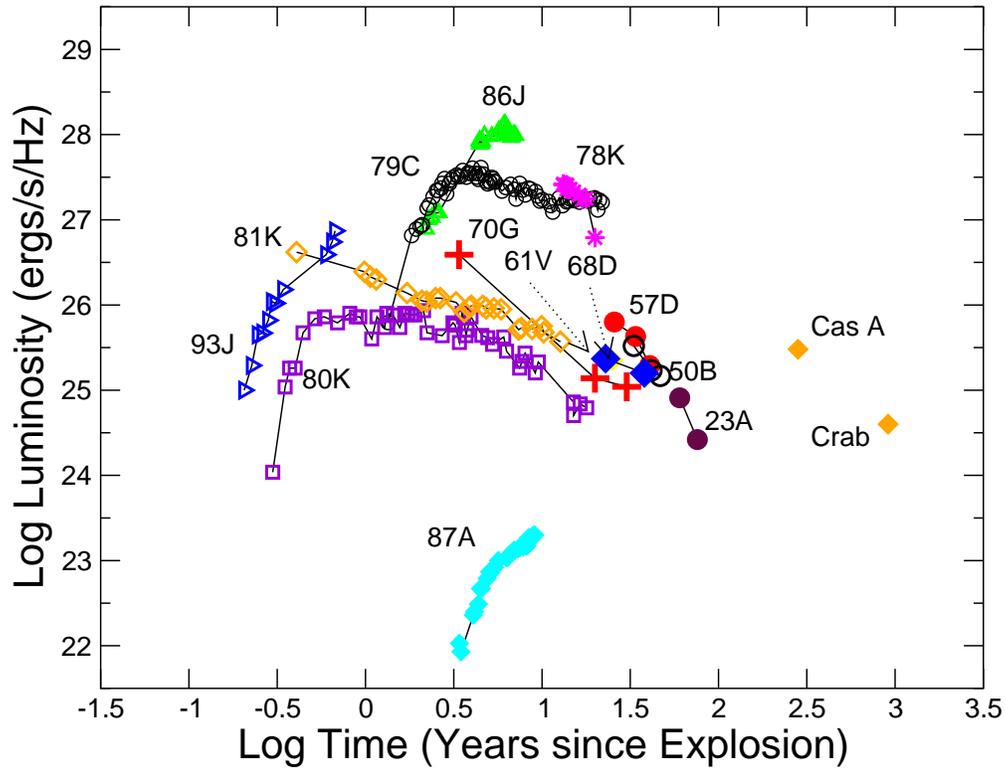}
\caption[Radio light curves for historical type II SNe in M83 at 20~cm                                
compared to several radio SNe II and SNRs.]{20~cm radio light curve  of several radio SNe II 
and SNRs.  Data and distances for SNe~1923A (maroon filled circle and maroon, open, inverted triangle for upper limit),             
1950B (black open circles) \& 1957D (red filled circles); for SN~1961V (blue filled diamonds); for SN~1968D (yellow filled diamond);                                       
 SN~1970G (red crosses); for SN~1978K (magenta stars);                                        
for SN~1979C (black open circles); for SN~1980K (purple open boxes); for SN~1981K (orange open diamonds); 
for SN~1986J (green triangles); for SN~1987A (cyan filled diamonds); and for SN~1993J 
(blue open triangles).  
Luminosities for Cas~A and the Crab (orange filled diamonds).
\label{lcfg}}

\end{figure}

The observed radio properties of SNe~1923A, 1950B, and 1957D are typical of values
reported for other radio SNe
(See Figure \ref{lcfg}).
It is clear, for example, from Figure \ref{lcfg} that the inferred radio luminosities
of SNe 1957D and 1950B are
(at a similar age of $\simeq$ 30$-$40 years) very similar to to the luminosities of
SN~1961V, in NGC~1058, and SN~1970G, in M101, at the same stage in their evolution. 
This correlation in luminosity also lends credence to our identification of
SNe~1950B as a Type II SN.  There is no reliable spectrum or optical
light curve of SN~1950B to make a definitive classification as a type II SN.  
In particular, there has
been some  uncertainty in
the optical position of SN~1950B that prevented a conclusive identification
of the supernovae with the
radio source \citep{cb85,crb94}.

The evolution of the radio flux density of SNe~1923A, 1950B, and 1957D
is also consistent with the current models for radio emission from SNe,
which predict a general decline in radio luminosity with age and
declining density of CSM.
Figure \ref{lcfg} illustrates that trend for a number 
of decades-old radio SNe,
with a gradual fading of the radio light curves 10 years after
the supernova event.  
Data, fits and distances for SNe~1923A, 1950B, and 1957D are taken from this paper and \citet{thi03};
for SN~1961V \citep{sto01a,sil96}; 
for SN~1968D \citep{hym95,tul88};                                       
for SN~1970G \citep{sto01b,cow91,kel96}; 
for SN~1978K, \citep{ryd93,sch99,tul88};      
for SN~1979C, \citep{wei86,wei91,mon00,fer96};
for SN~1980K, \citep{wei86,wei92,mon98,tul88};
for SN~1981K, \citep{van92,fre01};
for SN~1986J, \citep{rup87,wei90,sil96};
for SN~1987A, \citep{bal95,mit01};
for SN~1993J, \citep{van94,fre01}; and
for Cas~A and the Crab \citep{eck98}.

The differences in the  behavior of the individual SNe
are likely explained in terms of the density of the material
encountered by the supernova shock.
Thus, for example, 
the shocks associated with some radio SNe
(e.g. SNe~1979C, 1970G, 1961V, and early observations of SN~1950B) might be 
traveling through considerably denser CSM than
other similarly-aged radio SNe (e.g., SNe~1980K, 1978K, 1957D, and 1923A).
Consistent with this interpretation is the very rapid decline in
the radio emissions of Type Ib radio SNe, e.g. SNe~1983N and 1984L,
which presumably have lower density  CSM \citep{wei86,spw84,pan86}.  

One scenario which might explain
an increased density of CSM around some Type II radio SNe
could be that the progenitors underwent large-scale eruptions,
akin to luminous blue variables (LBVs), prior to the supernova event. 
The mass loss rates during an LBV eruption
can be 10---100$\times$ larger than the typical supergiant mass-loss rate
\citep{hum94}.  The exact epoch at which this may have occurred depends
on the ejection velocities of the CSM during these events, the rate
of expansion of the supernova shock, and the density of the CSM. 
Unfortunately these objects are too undersampled to make any definitive
statements as to the exact nature of such a possible outburst or mass loss.
Clearly, additional radio monitoring  of SNe~1957D,1950B, 1923A, and other
radio SNe, will be important in understanding 
the continuing evolution and nature of these
relatively rare objects.

The 1998 observations detected no radio emission from SNe 1945B, 1968L, and 1983N.
SNe~1945B and 1968L have never been detected in the radio. SN~1945B is located along a spiral arm and 
nothing can be inferred by its non-detection as very little is known about this SN \citep{lil90}. SN~1968L is located 
within a region of diffuse radio emission (background level $\simeq$ 3 mJy) in the nuclear region 
of M83.  SN~1983N was detected by \citet{cb85} within 1 year after discovery with flux 
densities at 20cm and 6cm of $4.1\pm0.1$ mJy and $0.70\pm0.06$ mJy, the
brightest historical radio SN detected in M83.  SN~1983N was likely a type Ib SN and is not expected to be a detectable radio source at this epoch as 
type I SNe have typically very low density CSM \citep{van96,wei02}.  SN~1968L would need to have been at least 3 times more luminous 15 years after explosion 
than SN~1983N (in the 1983 epoch) to have been detectable in the 1983 observations and in subsequent observations in the 1990's.  The diffuse nuclear radio emission coupled with the age of SN~1968L make it highly unlikely that it will ever be detected in the radio at the 1\arcsec \ resolution scale. 
Very long baseline interferometric (VLBI) observations have been made to study the nuclear region and
search for radio emission from SN~1968L using the Long Baseline Array of 
the Australia Telescope National Facility.  These results will be presented in a later publication.

\subsection{X-ray Constraints on Historical SNe}  

Using the deep Chandra observation, we can put tight constraints                          
on the X-ray emission and CSM properties of the historical SNe in M83.                          
Assuming a thermal plasma emission with a temperature of $kT = 0.8$~keV,                        
typical for the late emission originating in the reverse shock,                                 
the upper limits to the (0.3 -- 8~keV) X-ray luminosity are a few                                 
$\times 10^{36}~{\rm ergs~s}^{-1}$ ($3\sigma$; see Table~1). If the CSM                          
densities are dominated by the winds blown by the progenitor stars of the                       
SNe, we can use the X-ray interaction luminosity                                               
$L_{\rm x} = 4/[\pi m^2_{\rm H+He}) \Lambda(T) \times (\dot{M}/v_{\rm                           
w}]^2\times(v_{\rm s} t)^{-1}$                                                                  
at time $t$ after the outburst to measure the ratio $\dot{M}/v_{\rm w}$                         
\citep{imm03}. Assuming that $v_{\rm w}$ did not change                          
over time,                                                                                      
we can even directly estimate the mass loss rates of the progenitors.                           
%Integration of the mass-loss rate along the path of the expanding shell gives                                                                                       
%the mean density inside a sphere of radius $r$. For a constant wind velocity                                                                                        
%$v_{\rm w}$ and mass-loss rate $\dot{M}$ a         
%$\rho_{\rm csm} = \rho_0 (r/r_0)^{-s}$ profile with $s=2$ is expected.
                                                                                                
For each of the SNe we estimate mass-loss rates of                                              
$\dot{M} \approx 10^{-8}~M_{\odot}~{\rm yr}^{-1} (v_{\rm w}/10~{\rm                             
km~s}^{-1})$.                                                                                   
Progenitors of Type II SNe following core collapse of massive   
stars have high mass loss rates ($\dot{M} \sim   
10^{-4}$--$10^{-6}$~M$_{\odot}$~yr$^{-1}$)                   
and low wind velocities of typically $v_{\rm w}\sim10~{\rm km~s}^{-1}$.
Type Ib/c SNe have lower mass-loss rates                      
($\dot{M}\sim10^{-5}$--$10^{-7}~M_{\odot}~{\rm yr}^{-1}$) and 
significantly higher wind velocities of $10~{\rm km~s}^{-1} < v_{\rm w} \lesssim 1,000~{\rm km~s}^{-1}$.
                                                                                    
Our inferred mass-loss rates are consistent with previous observations of                       
other decades-old SNe. 
SN~1980K
was reported with an X-ray luminosity of $3.1\times10^{37}$ erg s$^{-1}$ measured with ROSAT in 1992 in the 0.1 -- 2.4~keV range and was not detected in Chandra observations in 2001 with a limiting detection threshold of $\sim 10^{37}$ erg s$^{-1}$ in the 0.3 -- 5~keV range \citep{sch94,hol03}.  There have been numerous X-ray SNe detected with ${\rm L_X}\sim 10^{38} $--$ 10^{41}$ erg s$^{-1}$ and there
has been a clear correlation between their X-ray and radio luminosities in the
first ten years following the SN explosion \citep{poo02}.  No X-ray emission has been detected from SNe with ages comparable to those of SNe~1957D, 1950B, and 1923A, 
with upper limits established for SNe~1959D of (in NGC~7331) of $1.2\times 10^{38}$ 
erg s$^{-1}$ and 1961V  (in NGC~1058) of $1.5\times 10^{40}$ erg s$^{-1}$
\citep{sto98b,sto98a}.
                                                               
Each X-ray observation at time $t$ is related to the corresponding                              
distance $r$ from the site of the explosion, $r=v_{\rm s} \times t$ (with                       
shock                                                                                           
front velocity $v_{\rm s}$), and to the age of the stellar wind,                                
$t_{\rm w} = t v_{\rm s}/v_{\rm w}$. The Chandra measurements                             
decades after the outbursts of the historical SNe in M83 probe the CSM at                       
large                                                                                           
radii ($>10^{18}$~cm) from the sites of the explosions. The observed lack                       
of significant X-ray emission is due to the lack of shocked CSM                                 
originating                                                                                     
in the low-density stellar winds of the progenitors.

\section{Summary}

M83 serves as an excellent laboratory to study decades-old SNe as they transition into SNRs.  The 15 year study 
by Cowan and collaborators has provided valuable insight into the evolution of these historical SNe and the nature
of the environment.  We report the continued decline in the non-thermal radio emission from SN~1957D and the
apparent fading of SN~1923A and 1950B below the confusion level of associated or intervening HII regions. 
We further report that SN~1983N was not detected in the latter epochs
since its initial discovery in 1983 and its behavior is typical of type Ib and Ic SNe.  The radio
non-detections of SNe~1945B and 1968L are also noted, although very little can be inferred from these
results due to lack of information about these sources and (in the case of SN~1968L) strong diffuse emission
in the nuclear region of M83.  Finally, we report the X-ray non-detection of all six historical SNe with Chandra.
These results place more stringent constraints on the mass-loss rates of the progenitors than was previously possible.
VLA observations are needed to explore the radio emission in these decades-old SNe and to continue to 
chart their evolution from SNe into SNRs.

\acknowledgments
We thank Michael P. Rupen for his assistance in imaging the 20~cm 1998 VLA data, Christina Lacey who
was the Principle Investigator for the 1998 radio observations, and
Emily Wolfing for her contributions through the NSF/REU program at the University of Oklahoma.  We also express
our appreciation to an anonymous referee, whose comments were very useful in improving this manuscript.
The research was supported in part by the
NSF (AST-0307279 to JJC), Chandra (GO1-2092B), the NASA Wisconsin Space Grant Consortium (CJS), and 
by an award from Research Corporation (C. Stockdale is a Cottrell Scholar).   
We have made use of the
NASA/IPAC Extragalactic Database (NED), which is operated by the Jet Propulsion Laboratory,
Caltech, under contract with the National Aeronautics and Space Administration, and have made use of 
the new NRAO Data Archival Service.

\clearpage
\pagestyle{empty}

\begin{deluxetable}{ccccccccccc}
%\begin{tabular}{ccccccccccc}
\rotate
\tablecolumns{11}
\tabletypesize{\small}
\tablewidth{0pt}
\tablecaption{
Radio flux densities and upper limits to the X-ray
luminosity for historical SNe in M83. Note: 1 mJy$=2.41\times 10^{25}$erg s$^{-1}$ Hz$^{-1}$
at a distance of 4.5 Mpc \citep{thi03}.  \tablenotemark{a}}
\tablehead{  
\colhead{} &\multicolumn{2}{c}{Position} &\multicolumn{3}{c}{1998 Flux Densities} &
\multicolumn{3}{c}{$1990-1992$\ Flux Densities} & \colhead{${\rm L_X}$ (0.3 -- 8~keV)} \\
\cline{2-9}
\colhead{Source}  & \colhead{RA}     & \colhead{Dec}    & \colhead{20cm} & \colhead{6cm} & \colhead{Spectral} & \colhead{20cm} 
& \colhead{6cm} & \colhead{Spectral} &
\colhead{$3\sigma$ upper limit} \\
  \colhead{SN}    & \colhead{(J2000)} & \colhead{(J2000)} & \colhead{(mJy)} & \colhead{(mJy)} &  \colhead{Index} & 
\colhead{(mJy)} & \colhead{(mJy)} & \colhead{Index} & \colhead{(ergs s$^{-1}$)} \\
}
\startdata
1957D &$13^{\mathrm{h}}37^{\mathrm{m}}03\fsec57$&$-29\deg49\amin40\fasec6$&$0.73\pm0.09$&$0.64\pm0.05$&$-0.11\pm0.15$&$1.75\pm0.07$ 
&$1.50\pm0.04$ &$-0.13\pm0.06$ & 2.2$\times10^{36}$\\
1950B &$13^{\mathrm{h}}36^{\mathrm{m}}52\fsec08$&$-29\deg51\amin55\fasec7$&$0.50\pm0.05$&$0.47\pm0.04$&$-0.05\pm0.13$ &$0.52\pm0.05$ 
&$0.49\pm0.04$ &$-0.05\pm0.13$ & 1.4$\times10^{36}$\\
1923A &$13^{\mathrm{h}}37^{\mathrm{m}}09\fsec31$&$-29\deg51\amin00\fasec7$&$\le0.15$&$0.21\pm0.04$&$\ge+0.28$ &$0.17\pm0.05$ 
&$0.21\pm0.03$ &$+0.18\pm0.33$ & 2.2$\times10^{36}$\\
1983N &$13^{\mathrm{h}}36^{\mathrm{m}}50\fsec31$&$-29\deg54\amin02\fasec3$&$\le 0.15$&$\le 0.12$& $\cdots$ &$\cdots$ &$\cdots$ &$\cdots$
& 1.7$\times10^{36}$\\
1945B &$13^{\mathrm{h}}36^{\mathrm{m}}51^{\mathrm{s}}\phantom{00}$ &$-30\deg10\amin17\asec\phantom{0}$&$\le 0.15$&$\le 0.12$& $\cdots$ &
$\cdots$&$\cdots$ &$\cdots$ & 3.6$\times10^{36}$\\
\enddata
%\end{tabular}
%\caption{Radio flux densities (obtained in 1998) and upper limits to the X-ray
%luminosity for historical SNe in M83. Note: 1 mJy$=2.41\times 10^{25}$erg s$^{-1}$ Hz$^{-1}$ at  distance of 4.5 Mpc
%\citep{thi03}.
%SN 1968L is located in the confused nuclear region of the galaxy.  We were unable to get limits on either radio or X-ray luminosities.
\tablenotetext{a}{SN~1968L is located in the confused nuclear region of the galaxy.  
We were unable to get meaningful limits for either radio or X-ray luminosities.}
\end{deluxetable}

\clearpage
\pagestyle{empty}

\end{document}